\documentclass[journal]{IEEEtran}
\usepackage{amsmath}
\usepackage{graphicx}
\usepackage{color} 
\newcommand{\comment}[1]{} 
\begin{document}

\title{Detection of Anomalous Reactor Activity Using Antineutrino Count Rate Evolution Over the Course of a Reactor Cycle}
\author{ Vera Bulaevskaya, Adam Bernstein}
\maketitle

\begin{abstract}
This paper analyzes the sensitivity of antineutrino count rate measurements to changes in the fissile content of civil power reactors. Such measurements may be useful in IAEA reactor safeguards applications. We introduce a hypothesis testing procedure to identify  statistically significant differences between the antineutrino count rate evolution of a standard 'baseline' fuel cycle and that of an anomalous cycle, in which plutonium is removed and replaced with an equivalent fissile worth of uranium. The test would allow an inspector to detect anomalous reactor activity, or to positively confirm that the reactor is operating in a manner consistent with its declared fuel inventory and power level. We show that with a reasonable choice of detector parameters, the test can detect replacement of 73 kg of plutonium in 90 days with 95\% probability, while controlling the false positive rate at 5\%. We show that some improvement on this level of sensitivity may be expected by various means, including use of the method in conjunction with existing reactor safeguards methods. We also identify a necessary and sufficient daily antineutrino count rate to achieve the quoted sensitivity, and list examples of detectors in which such rates have been attained.
\end{abstract}

\begin{IEEEkeywords}
Safeguards applications, neutrinos, nuclear monitoring, nuclear power plants, nuclear power simulation, particle detectors 
\end{IEEEkeywords}

\section{Introduction}
\label{intro}
The International Atomic Energy Agency (IAEA) nuclear safeguards regime is designed to detect diversion of fissile material from civil nuclear fuel cycle facilities to weapons programs \cite{IAEAsg}. In previous work, we  predicted \cite{SONGS1} and demonstrated \cite{nuresults},\cite{nupower},\cite{nuburnup} that cubic meter scale antineutrino detectors, operating at a distance of tens of meters from a 1 gigawatt electric (GWe) pressurized water reactor (PWR), can directly detect changes in operational status, power levels, and fissile inventory of the reactor core.  Similar results were achieved earlier by a Russian group \cite{Klimov}. These metrics are all of potential use for the IAEA reactor safeguards regime. 

In this paper, we demonstrate a possible methodology for using antineutrino detection in a safeguards context. We introduce a hypothesis testing procedure to identify statistically significant differences between the antineutrino count rate evolution of a standard 'baseline' fuel cycle and that of an anomalous cycle in which  73 kg of plutonium has been removed and replaced with the equivalent fissile worth of uranium. (This quantity of plutonium represents the removal and replacement of ten partially burnt assemblies with ten fresh fuel assemblies.) The test would allow an inspector to detect  anomalous reactor activity, or to positively confirm that the reactor is operating in a manner consistent with its declared fuel inventory and power level. We show that with a reasonable choice of detector parameters, the test can detect the net loss from the core of 73 kg of plutonium  in 90 days with 95\% probability, while controlling the false positive rate at 5\%.  

The purpose of the study is to explore this possible alternative method of reactor safeguards, by quantifying the sensitivity of an antineutrino count rate measurement to anomalous changes in fissile content. In describing our example, we avoid the standard IAEA term 'diversion', since we do not explicitly specify the fate of the removed plutonium. In particular, we are not asserting that the removal of plutonium in this example could not be uncovered by existing IAEA safeguards methodologies. 

One of IAEA's inspection goals is to be able to detect diversion of 8 kg \footnote{8 kg is   designated by the IAEA as a  'significant quantity'  (SQ) of plutonium. The IAEA definition of a significant quantity is 'the approximate quantity of nuclear material in respect of which, taking into account any conversion process involved, the possibility of manufacturing a nuclear explosive device cannot be excluded' \cite{IAEAaccounting}.} of plutonium  from a civil nuclear facility in a 90-day period \cite{harms-rod}. Our current sensitivity to anomalous reactor operation caused by removal of plutonium is at the level of several significant quantities.  Enhancements to the detector, including the capability to measure the antineutrino energy spectrum, may allow for detection of even smaller changes in the reactor's fissile content. While the demonstrated sensitivity is not to actual diversion but to anomalous reactor operations, we expect that this method can be used in conjunction with existing IAEA safeguards methodologies to achieve IAEA SQ goals for diverted material. We note that  other IAEA surveillance and accountancy measurement devices do not in isolation reach the SQ goals, but are used as part of a comprehensive accountancy strategy. Examples include Cherenkov light monitors in spent fuel cooling ponds, which are not sensitive at the SQ level, but which provide continuity of knowledge and confirm the presence of large numbers of radioactive spent fuel assemblies. 

We begin by briefly describing the relationship between the antineutrino count rate and the reactor fissile  inventory, and contrast our method for anomaly detection with current IAEA reactor safeguards practice. Next, we describe the test procedure and its inputs, including the fuel loadings of the baseline and anomalous scenario cycles. We then examine the statistical power of the procedure to distinguish between the two cycles and thereby identify an anomaly in reactor operations. We include the effects of counting statistics, a fixed systematic bias in detector response,  deliberate malfeasance on the part of the reactor operator, the starting point and duration of data acquisition, and simulation errors.  We also establish a range of acceptable detector masses, intrinsic efficiencies and standoff distances that would permit discovery of the anomaly in our example.  We conclude by summarizing the potential impact of this approach on current IAEA safeguards and useful next steps. 
\par
In this paper, we assume that the background count rates are negligible relative to those produced by the antineutrino signal. This assumption is based on high signal-to-background ratios achieved in several past experiments, discussed in Section \ref{detector}. If these ratios are not achieved, the results described here do not apply, and an additional analysis would be required to account for the effects of higher background rates.

\section{ Current IAEA Reactor Safeguards and Antineutrino-Based Safeguards} 
\label{current}

Currently, the IAEA uses nuclear material accountancy, as well as containment and surveillance (C\/S) techniques to verify the quantities of fuel used in and discharged from reactors.  Nuclear material accountancy refers to a quantitative and independent check of  fuel inventories, performed by the Agency. At reactors, the predominant material accountancy method is item accountancy, or counting of items (fresh and spent fuel assemblies and rods) considered to contain fixed and known quantities of fissile material.  The presence and integrity of radioactive spent fuel assemblies and rods in cooling ponds at the reactor is also checked by Cherenkov light measurements and other methods.  C\/S techniques, such as videocameras and seals on the reactor head, are also used \cite{IAEAaccounting}. 

By contrast, antineutrino-based safeguards offer a form of near-real-time and nondestructive bulk accountancy. In contrast to item accountancy, bulk accountancy methods provide estimates of the total fissile mass without relying on assumptions about the mass contents of premeasured items. Examples include coincidence neutron counting, mass spectroscopy and chemical analyses.  As such, antineutrino based methods are complementary to the existing safeguards regime, since they provide independent quantitative information about fissile material inventories as long as the reactor is operational. Among other uses, this information can provide independent confirmation that the fuel inventory at beginning and throughout the reactor cycle is consistent with operator declarations. In principle, the inventory estimate so derived can also be used to check for shipper/receiver differences, both for fresh fuel taken in by the operator and for spent fuel sent to downstream reprocessing or storage facilities.

While the measurement capability appears promising, its actual import for IAEA safeguards is beyond the scope of this paper.  As an example of the complications that arise, we note that for existing power reactors, the antineutrino-based inventory estimates would have to be reconciled with and integrated into the full accounting of all materials at the reactor site, including that in  spent fuel cooling ponds. For such sites, with decades of accumulated and largely unassayed fuel, containing many tens of tons of  fissile material, such accounting may prove impractical. For this reason, we recommend that a more detailed analysis of the capability be conducted by safeguards experts,  both for existing and future reactor safeguards regimes.

\section{Modeling the Antineutrino Count Rate for Safeguards Applications} 
\label{modelintro}

 A change in fissile mass content in a reactor core -  such as that occurring when uranium is consumed and plutonium produced in the course of a reactor fuel cycle -  creates a measurable systematic shift in the antineutrino count rate (and energy spectrum). In previous work \cite{nuburnup}, we have shown that the antineutrino count rate is reduced by about 10$\%$ relative to its value at the beginning of the cycle over the course of a typical 1.5 year pressurized water reactor (PWR) fuel cycle. This reduction occurs even when (as is typical) the reactor maintains constant power throughout the cycle; therefore, monitoring the antineutrino count rate provides information about core fissile inventory evolution that is not accessible through a measurement of the reactor power alone. 

In a safeguards context, the measured antineutrino count rate evolution would be compared to a predicted count rate evolution assuming normal conditions (i.e., no removal of plutonium) over some portion or all of the fuel cycle. The predicted evolution under normal operating conditions will be referred to as the ``baseline scenario'' for the remainder of this paper.  The prediction is obtained from a reactor simulation code, such as ORIGEN \cite{ORIGEN}, which takes as inputs the operator-declared thermal power and initial fissile isotopic masses, as well as other reactor parameters, and returns fission rates for each isotope. The individual fission rates are then converted into a predicted emitted antineutrino flux using standard analytical formulae.  The emitted antineutrino flux is finally converted to a measured antineutrino count rate, using a detector response function derived from experiment and modeling. 

In the present work, we simulate both the baseline and anomalous antineutrino count rates over the course of the fuel cycle for use in our hypothesis test. We use an ORIGEN simulation of the core of Unit 2 of the San Onofre Nuclear Generating Station (SONGS), originally published in \cite{misnerthesis}. The detector response function was derived from the SONGS1 experiment \cite{SONGS1}, for which the  antineutrino signal was approximately 360 counts per day at beginning of cycle after subtraction of reactor-off backgrounds. 

Following \cite{Klimov}, we describe the PWR core antineutrino count rate evolution $N_{\bar{\nu}}(t)$ at time $t$ in the fuel cycle as a product of two time-dependent factors:
\begin{equation}
\label{kgamma}
 N_{\bar{\nu}}(t) =P_{th}(t) \cdot \gamma (1+k(t)).
\end{equation}   
$P_{th}(t)$ is the reactor thermal power. The term $(1 + k(t))$ depends on the changing fissile isotopic content of the core, embodied in the parameter $k(t)$. $\gamma$ is a constant related to the detector mass, efficiency, and standoff distance. This parametrization highlights the direct  dependence of the count rate on the thermal power, an important consideration we return to in Section \ref{malfeas}.

For the PWR core being considered here, (\ref{kgamma}) is well approximated by a quadratic function of time: 
\begin{equation}
\label{model}
N_{\bar{\nu}}(t) = \beta_0 + \beta_1t + \beta_2t^2.
\end{equation}   
The quadratic model in (\ref{model}) is valid for PWRs loaded with typical Low Enriched Uranium (LEU) fuel. Other fuel loadings and reactor types can result in an antineutrino count rate evolution that is substantially different in form from (\ref{model}). 
\par
The coefficients $\beta_0$, $\beta_1$ and $\beta_2$ in (\ref{model}) can be used to detect a departure from the baseline scenario. The measured antineutrino count rate evolution can be used to estimate the coefficients, which can then be compared to those predicted for the baseline scenario. A statistically significant difference in at least one of the estimated coefficients from its baseline counterpart could indicate a departure of the observed evolution from that of the baseline scenario.

\section{Testing For Anomalous Activity}
\label{testproc}
Following the model in (\ref{model}), the true baseline evolution of antineutrino count rate as a function of time $t$ in the fuel cycle is given by
\begin{equation}
\label{basemodel}
N^{(B)}_{\bar{\nu}}(t) = \beta^{(B)}_0 + \beta^{(B)}_1t + \beta^{(B)}_2t^2 
\end{equation}
(The superscript {\it ``B''} in the above equation and for the remainder of the paper indicates ``baseline'').  As discussed earlier, the true baseline evolution is obtained from a reactor simulation. To account for simulation error, we modify the model in (\ref{basemodel}) by representing the baseline count rate at time $t$ as Gaussian with the mean equal to the simulation value and the standard deviation equal to 1\% of the simulation value, i.e.,  
\begin{equation}
\label{basemodel.gauss}
N^{(B)}_{\bar{\nu}}(t) \sim \ \mbox{Gaussian}(\mu(t), 0.01\mu(t)). 
\end{equation}
$\mu(t)$ is the baseline evolution antineutrino count rate value at time $t$ from the simulation and can be modeled as 
\begin{equation}
\label{mu.beta}
\mu(t) = \beta^{(B)}_0 + \beta^{(B)}_1t + \beta^{(B)}_2t^2.
\end{equation}
One percent random error is typical for these and other ORIGEN simulations \cite{misnerthesis},\cite{ORIGENbenchmarks}. (Systematic shifts of the predicted and measured response are treated separately in Section \ref{detectbias}.)
\par
Let $\{N^{(M)}_{\bar{\nu}}(t)\}$ denote the measured count rate evolution (the superscript {\it ``M''} indicates ``measured'') which is to be tested against the baseline scenario evolution. Since the measurements follow Poisson statistics, 
\begin{equation}
\label{testmodel}
N^{(M)}_{\bar{\nu}}(t) \sim \ \mbox{Poisson}(\beta^{(M)}_0 + \beta^{(M)}_1t + \beta^{(M)}_2t^2). 
\end{equation}
To determine whether the measured antineutrino count rate evolution deviates significantly from that of the baseline, we can compare the coefficient $\beta^{(B)}_i$ in (\ref{mu.beta}) to its counterpart $\beta^{(M)}_i$ in (\ref{testmodel}) for each $i = 0, 1, 2$.  This requires us to estimate each of these coefficients.
\par
 One way to do this is to perform the least squares (LS) regression of both the modeled baseline count rates $N^{(B)}_{\bar{\nu}}(t)$ and the measured count rates $N^{(M)}_{\bar{\nu}}(t)$ on $t$ and $t^2$.  LS regression is best suited to the case of Gaussian noise with constant variance \cite{neter-wasser-kutner}.  In our case, the baseline count rates do in fact have Gaussian noise by construction, and high Poisson statistics make the noise in the measured count rates approximately Gaussian.  Moreover, as noted in Section \ref{modelintro}, the change in the count rate variance over the course of the cycle for a standard PWR is about 10\%.  Thus, LS regression should produce statistically near-optimal coefficient estimates in this context (if necessary, weighted least squares regression could be used to alleviate the issue of non-constant variance).    
\par
Of greater concern for the regression analysis is that $t$ and $t^2$ are highly correlated, which can lead to very unstable coefficient estimates.  A common way to overcome this problem is to perform regression on deviations from the sample mean of times $(t - \bar{t})$ and deviations from the mean squared $(t - \bar{t})^2$ because the correlation between these two terms is substantially lower than that between $t$ and $t^2$ \cite{neter-wasser-kutner}. Therefore, we reparameterize the model for the measured count rates $N^{(M)}_{\bar{\nu}}(t)$ in (\ref{testmodel}) as follows:
\begin{equation}
\label{testmodel1}
N^{(M)}_{\bar{\nu}}(t) \sim \ \mbox{Poisson}(\gamma^{(M)}_0 + \gamma^{(M)}_1(t-\bar{t}) + \gamma^{(M)}_2(t-\bar{t})^2). 
\end{equation}
We must also reparametrize the  model in (\ref{mu.beta}). The baseline count rate $N^{(B)}_{\bar{\nu}}(t)$ still follows (\ref{basemodel.gauss}), but the baseline mean function $\mu(t)$ is now given by 
\begin{equation}
 \label{mu.gamma}
\mu(t) = \gamma^{(B)}_0 + \gamma^{(B)}_1(t-\bar{t}) + \gamma^{(B)}_2(t - \bar{t})^2.
\end{equation}
Each coefficient $\gamma^{(M)}_i$ in (\ref{testmodel1}) can then be compared to its counterpart $\gamma^{(B)}_i$ in (\ref{mu.gamma}) for $i = 0, 1, 2$ by testing the following pairs of hypotheses:
\begin{eqnarray}
\label{hyp.gamma}
H^{(0)}_o:  \gamma^{(M)}_0 = \gamma^{(B)}_0 \ & \mbox{versus}  \ & H^{(0)}_a:  \gamma^{(M)}_0 \neq \gamma^{(B)}_0     \nonumber \\
H^{(1)}_o:  \gamma^{(M)}_1 = \gamma^{(B)}_1 \ & \mbox{versus}  \ & H^{(1)}_a:  \gamma^{(M)}_1 \neq \gamma^{(B)}_1     \\
H^{(2)}_o:  \gamma^{(M)}_2 = \gamma^{(B)}_2 \ & \mbox{versus}  \ & H^{(2)}_a:  \gamma^{(M)}_2 \neq \gamma^{(B)}_2     \nonumber 
\end{eqnarray}

\par
The test procedure then consists of the following steps:
\begin{enumerate}
\item Generate $\{N^{(B)}_{\bar{\nu}}(t)\}$ according to (\ref{basemodel.gauss}).  
\item  Obtain coefficient estimates $\hat{\gamma}^{(B)}_0$, $\hat{\gamma}^{(B)}_1$ and $\hat{\gamma}^{(B)}_2$, and their standard errors $se(\hat{\gamma}^{(B)}_0)$,  $se(\hat{\gamma}^{(B)}_1)$ and  $se(\hat{\gamma}^{(B)}_2)$ from the least squares regression of generated count rates $\{N^{(B)}_{\bar{\nu}}(t)\}$ on time deviations $(t - \bar{t})$ and time deviations squared $(t - \bar{t})^2$ (where $\bar{t}$ is the sample average of all the time values $t$).   
\item Similarly, obtain coefficient estimates $\hat{\gamma}^{(M)}_0$, $\hat{\gamma}^{(M)}_1$ and $\hat{\gamma}^{(M)}_2$, and their standard errors $se(\hat{\gamma}^{(M)}_0)$,  $se(\hat{\gamma}^{(M)}_1)$ and  $se(\hat{\gamma}^{(M)}_2)$ from the least squares regression of measured count rates $\{N^{(M)}_{\bar{\nu}}(t)\}$ on $(t - \bar{t})$ and $(t - \bar{t})^2$. 
\item Obtain test statistics
\begin{equation}
s_i = \frac{\hat{\gamma}^{(M)}_i - \hat{\gamma}^{(B)}_i}{\sqrt{se^2(\hat{\gamma}^{(M)}_i) + se^2(\hat{\gamma}^{(B)}_i)} }
\end{equation}
for $i = 0, 1, 2$ and their corresponding $p$-values, given by
\begin{equation}
\label{pval}
p_i = 2\cdot P(S>|s_i|) 
\end{equation}
where $S$ has a Student's $t$ distribution with $2\cdot (n - 3)$ degrees of freedom with $n$ equal to the number of count rate measurements.   
\item Determine the acceptable false positive (FP) rate (see Section \ref{testperf}) and apply the false discovery rate (FDR) procedure, described in \cite{benj-hoch}\footnote{As described in more detail in \cite{benj-hoch}, the FDR procedure controls the error rate of testing multiple hypotheses.}, to determine whether to reject each of the $H^{(i)}_o$ in favor of $H^{(i)}_a$ in (\ref{hyp.gamma}).  If at least one of the null hypotheses is rejected, conclude that the measured evolution deviates significantly from that of the baseline. Otherwise, conclude that the measured evolution does not significantly deviate from that of the baseline. 
\end{enumerate}

\section{Test Performance}
\label{testperf}
The test can produce two types of errors: it could find a significant difference from the baseline in at least one  coefficient  when the evolution was in fact produced by a baseline scenario (a false positive, or FP, result), or it could miss a significant difference in all  coefficients when the evolution was different from the baseline  (a false negative, or FN, result).  
\par
The complement of the FN rate is the true positive (TP) rate. The TP rate is defined as the probability of finding a significant difference in at least one of the coefficients from its baseline counterpart when the evolution in question is in fact different from that of the baseline. A good test has a low FP rate and a high TP rate. There is a trade-off between these two quantities:  all else being equal, increasing the TP rate of the test comes at a price of a higher FP rate. A Receiver Operating Characteristic (ROC) curve for a particular test procedure shows the former as a function of the latter, thus allowing one to determine the minimum FP rate that yields the desired TP rate. 

\subsection{ROC Curve Simulation}
\label{rocsim}
To estimate the ROC curve of the test, we carried out a simulation (not to be confused with the reactor simulation) that estimates the TP rate of the test for a given FP rate.  This simulation was performed for a scenario in which ten once-burned assemblies with the highest plutonium content  are removed and replaced with 3.91\% enriched fresh fuel. This represents the removal of 73 kg of $^{239}$Pu from the core. Complete fissile inventories at beginning of cycle for the baseline and anomalous scenario are shown in Table~\ref{fis_table}. 

\begin{table}[h!b!p!]\footnotesize
\centering
\caption{The initial inventories of the main fissioning isotopes for the baseline and anomalous scenarios. The final column is the difference in fissile content between the two scenarios.  A negative  (positive) value indicates that the isotope was removed (added) in the anomalous scenario.}
\begin{tabular}{|l||c|c||c|}
\hline\ \ 
         Isotope                       & Baseline mass   & Anomalous scenario  &  Mass difference  \\ 
                                & 			    (kg)  &  mass (kg)  			    &    (kg)                     \\ \hline
{\bf $^{235}$U}        & 2834                     & 2849                                 &  15  \\ \hline
{\bf $^{238}$U}         & 82912                   & 83351 				&  439  \\ \hline
{\bf $^{239}$Pu}       & 225                       &152				 & -73  \\ \hline
{\bf $^{241}$Pu}        & 21         		& 12 					& -9  \\ \hline

\end{tabular}
\label{fis_table}
\end{table}

\begin{figure}[!t]
\centering
\includegraphics[width=3.5in]{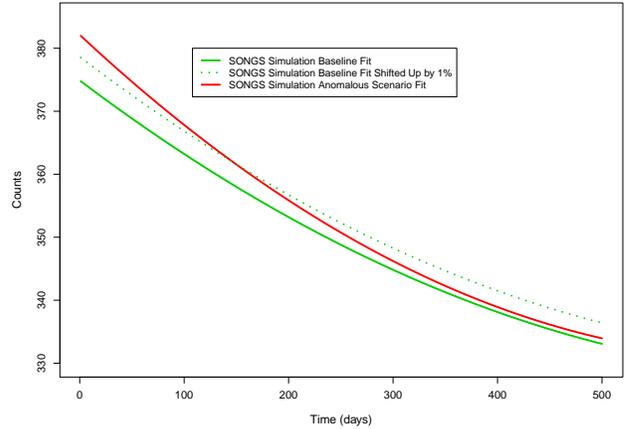}
\caption{Baseline (solid green), shifted baseline (dashed green) and anomalous  (red) scenario evolutions of daily antineutrino count rates versus time (in days), as simulated in ORIGEN.}
\label{counts_opershift}
\end{figure}

 Fig. \ref{counts_opershift} shows the antineutrino count rate evolutions predicted by the ORIGEN simulation for the baseline scenario (solid green) and the anomalous scenario (red). (The shifted baseline evolution, shown in dashed green,  is discussed in Section \ref{malfeas}).

A given point on a ROC curve is obtained as follows. One hundred thousand pairs of anomalous and baseline evolutions are generated, with the former from a Poisson distribution with the coefficients $\gamma^{(M)}_i$, $i = 0, 1, 2$, obtained from the ORIGEN reactor simulation for the given scenario and time period, and the latter from a Gaussian distribution according to (\ref{basemodel.gauss}).  The test procedure introduced in Section \ref{testproc} is then applied  at the given FP rate (the $x$ coordinate of the point on the ROC curve) to each pair of evolutions.  We then estimate the TP rate (the $y$ coordinate of the point on the ROC curve) with the fraction of the 100,000 evolution pairs for which at least one of the null hypotheses in (\ref{hyp.gamma}) is rejected.  This is repeated for a sequence of FP rate values from 0 to 1, thus producing a curve. The large number of generated evolutions ensures that every TP rate estimate is within 1\% of the relevant true TP rate. 
\par
To verify that the nominal FP rate of our test procedure corresponds to its actual FP rate, we also generated 100,000 baseline evolutions from a Poisson distribution with the coefficients $\gamma^{(B)}_i$, $i = 0, 1, 2$, obtained from the ORIGEN reactor simulation, for the given time period.  We estimated the actual FP rates with the fractions of these evolutions for which at least one of the null hypotheses in (\ref{hyp.gamma}) was rejected and found them to be very close to the nominal FP rates.  
\par
While the  performance of the test will depend on the specific scenario, the present example allows us to identify several important factors that influence our ability to detect any anomalous reactor operation.  In the following sections we assess the impact on our test performance of finite counting statistics, systematic error in the detector response, operator malfeasance, and the starting point and duration of data acquisition within the cycle.

\subsection{Effect of Counting Statistics}
\label{countstats}
For the evolutions shown in Fig. \ref{counts_opershift}, antineutrino count rates range from approximately 375 per day at the beginning of cycle to approximately 335 per day at the end of cycle.  As discussed in Section \ref{detector}, easily achievable increases in the combined detector mass and efficiency can lead to a five-fold improvement in counting statistics. We considered the impact of these changes on the test performance, simply by increasing the count rate used in our test by a factor of 5. 
\par
\begin{figure}[!t]
\centering
\includegraphics[width=3.5in]{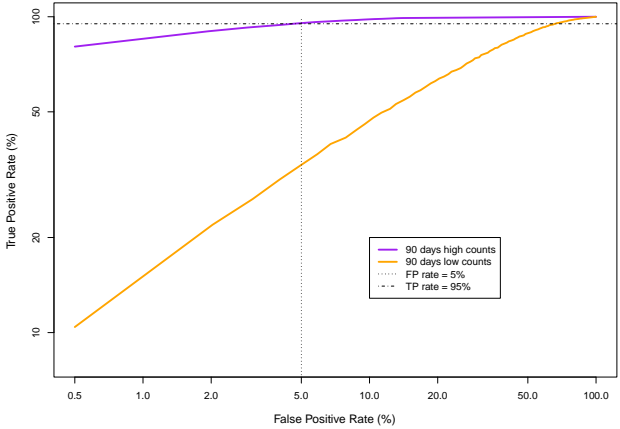}
\caption{ROC curves for the test using days 0--90, assuming low counts (orange) and high counts (purple). The dotted vertical line corresponds to the FP rate of 5\%, while the dashed and dotted horizontal line corresponds to the TP rate of 95\%.}
\label{roc_counts}
\end{figure}

Fig. \ref{roc_counts} shows that this dramatically improves the performance of the test.  The ROC curve for high count rates collected over the first 90 days in the cycle, shown in purple, is up to six times higher than the ROC curve for the low count rates for the same time period, shown in orange.  For example, at the FP rate of 5\%, the high count TP rate of the former is 95\%, while the low count TP rate is 34\%. This strong effect was observed for other data acquisition periods. These results, as well as those discussed in Sections \ref{malfeas} and \ref{duration}, are summarized in Table \ref{fp5perctable}. For the particular scenario considered here, we verified that a minimum five-fold improvement in counting statistics is necessary in order to achieve a 95\% TP rate at the 5\% FP rate. This was accomplished by progressively increasing the count rate in the testing procedure until the 95\% TP\% / 5\% FP standard was achieved.       
\begin{table}[h!b!p!]\footnotesize
\centering
\caption{True positive rates (in \%) at the false positive rate of 5\% for the various factors considered here. See text for an explanation of each factor.}
\begin{tabular}{|l||c||c||c|c||c|}
\hline
            & \multicolumn{4}{c||} {{\bf 2000 counts}} & {\bf 375 counts}\\ 
{\bf Dura-} &  \multicolumn{4}{c||} {{\bf per day}} & {\bf per day }\\ 
\cline{2-6} 
  \ {\bf tion}  & {\bf Un-} & \multicolumn{3}{c||} {\bf Shifted Due To} & {\bf Un-} \\ \cline{3-5}
    {\bf (days)}        & {\bf shifted}          &        {\bf Mal-}          & \multicolumn{2}{c||} {\bf Detector Bias}   & {\bf shifted} \\ \cline{4-5}
               &       & {\bf fesance}    & {\bf Uncorr.} & {\bf Corrected} & \\ \hline
first 30       & 58    &    & & & \\ \hline
first 90       & 95    & 23 & 0.4 & 12 & 34 \\ \hline
first 180      &       &    &     & 62 & \\ \hline
first 250      & 99    & 56 &     & 96 &  \\ \hline
last 90        &       & 32 &     &    & \\ \hline
last 250       &       & 73 &     &    & \\ \hline
500            & 99.99 & 99 &     &    & \\ \hline
\end{tabular}
\label{fp5perctable}
\end{table}

\subsection{Effect of a Systematic Shift in Detector Response}
\label{detectbias}
Systematic shifts in the detector response could cause upward or downward shifts in the measured antineutrino count rate.  In that case, even if fuel has not been removed, the detector measurements may deviate significantly from the predicted baseline evolution. In this section we analyze the consequences of such shifts for the hypothesis testing procedure.
\par 
The absolute count rate of reactor antineutrinos has been measured with 3$\%$ systematic uncertainty \cite{bemporad}. Antineutrino count rate measurements made relative to an initial value have a considerably smaller systematic error, of less than 1$\%$ \cite{dcloi}, since  fixed systematic errors  present in the absolute measurement are cancelled by subtraction. As we will show, a hypothesis test that uses antineutrino count rate trajectories made relative to a premeasured value are much less sensitive to systematic detector shifts than a test on data not referred to an initial value. 
\par
 In an actual safeguards deployment,  a detector bias would become evident by a comparison of  measured and predicted antineutrino counts integrated over a few weeks.  For example, with measured antineutrino count rates of  2000 counts per day, 20 days of data acquisition would suffice to reduce the statistical error to 0.5\%, small enough to measure a few percent difference between predicted and actual rates. In the context of the hypothesis test considered here, such a shift can be mistakenly interpreted as evidence for anomalous reactor operations, or correctly as a previously undiscovered systematic shift in the detector response, not attributable to the anomaly. 

\begin{figure}[!t]
\centering
\includegraphics[width=3.5in]{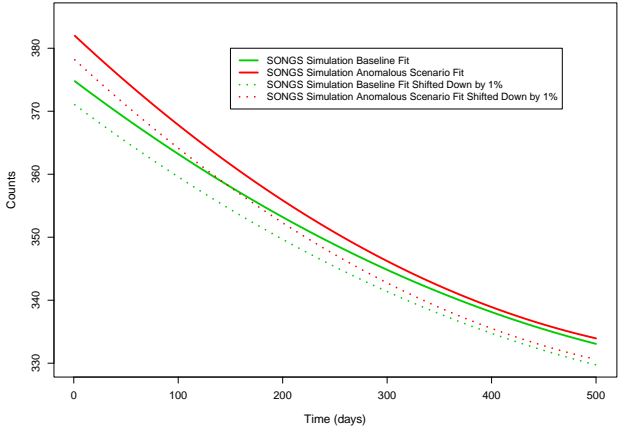}
\caption{Baseline (solid green),  anomalous (solid red), shifted baseline (dashed green) and shifted anomalous (dashed red) scenario evolutions of antineutrino daily count rates versus time (in days), as simulated in ORIGEN.}
\label{counts_detshiftneg}
\end{figure}

We examined the impact of a systematic shift incorrectly interpreted as evidence for anomalous reactor operations.  We adjusted both the baseline and anomalous measured count rate evolutions  by 1\%.  (We report only a downward shift result, which is conservatively worse than the impact of the upward shift for the scenarios considered here).  A 1\% absolute systematic error is smaller than that typically obtained in reactor antineutrino experiments, but is already large enough to illustrate the strong impact of detector bias.  Fig. \ref{counts_detshiftneg} shows the resulting shifted baseline and anomalous count rate evolutions, as well as the original unshifted evolutions. As can be seen from this plot, the shifted baseline evolution is now further from the reference (original) baseline than the shifted anomalous evolution. As a result, the performance of a test deteriorates dramatically. At 5\% FP rate, the TP rate is 0.4\%, compared to 95\% in the absence of a detector bias. The test attains the desired 95\% TP rate only at the FP rate of practically 100\%.  Thus, even a small bias in the detector response  severely weakens the statistical power of the hypothesis test if an absolute comparison of count rate trajectories is made. 
\par

The negative impact on the test of an absolute systematic shift in detector response can be mitigated in two ways: either by using relative count rate data, referred to a corrected value measured at startup, or by comparison with a template from a previous cycle known by other means to be standard.
\par
For the first case, we investigated the TP rate of the hypothesis test assuming  the measured antineutrino count rates are corrected by the difference between the predicted and measured values averaged over the first 20 days of the cycle. Thus, agreement of predicted and measured count rates at beginning of cycle is enforced before the testing procedure is applied. (This is equivalent to making an initial assumption that no anomaly is present. When the shifted antineutrino counts are corrected in this way, the TP rate is 12\% at 5\% FP rate with 90 days of data acquisition, which is a significant improvement over 0.4\% TP rate reported above in the case of shifted measurements not referred to an initial value (referred to as ``uncorrected'' in Table \ref{fp5perctable}).  When the acquisition period is increased to 180 days, this correction yields a TP rate of 62\% at the 5\% FP rate.  For 250 days, the TP rate is 96\%, which is only slightly below the rate in the absence of a shift for the same acquisition period.

While a measurement relative to startup improves the power of the test, the most favorable approach is to use a measured template for the antineutrino count rate, derived from a previous cycle known to be standard by other means.  By definition, this removes any systematic detector bias, since the relation between the baseline fuel evolution and the measured antineutrino count rate has been empirically established.  This case reverts to our earlier result for high statistics acquisition -  95\% TP rate and a 5\% FP rate with 90 days of data acquisition.
The approach of using a predefined template from a previous and well known fuel cycle has a further advantage that it no longer depends on a reactor simulation and its associated errors.  This appears to be the most effective method for identifying anomalous fuel loadings, so long as systematic errors in antineutrino detector predicted and measured response remain at the level of a few percent.

\par

\begin{figure}[!t]
\centering
\includegraphics[width=3.5in]{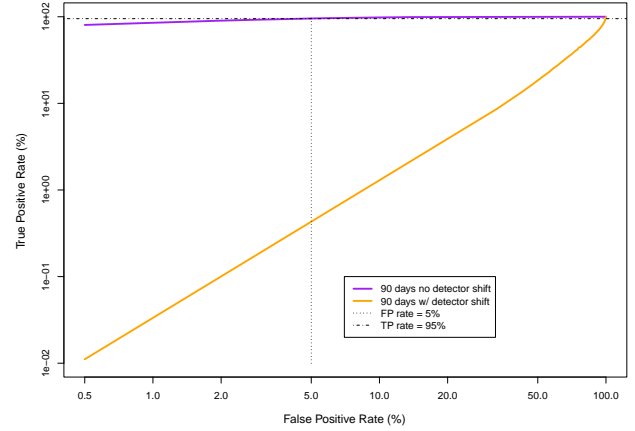}
\caption{ROC curves for the test applied to the original unshifted (purple) and detector-shifted (orange) evolutions.  The time period shown here is days 0-90 in the fuel cycle.  The dotted vertical line corresponds to the FP rate of 5\%, while the dashed and dotted horizontal line corresponds to the TP rate of 95\%.}
\label{roc_detshiftneg}
\end{figure}

\subsection{Effect of Operator Malfeasance}
\label{malfeas}
Equation (\ref{kgamma}) shows that both thermal power and fissile isotopic content can be altered to change the antineutrino count rate. Thus, in an attempt to conceal the removal of plutonium in the present example, the reactor operator could report a higher thermal power value than the true operating power. This input information would cause the simulation to  incorrectly predict a systematic upward shift in the baseline evolution. 
\par
To assess the impact of a misreported power history, we considered the effect of a 1\% upward systematic shift of the baseline evolution that was originally obtained from the ORIGEN simulation (solid green curve in Fig. \ref{counts_opershift}).  Fig. \ref{counts_opershift} shows the resulting shifted baseline evolution (dashed green curve).  As can be seen from the plot, this evolution is much less distinguishable from the anomalous evolution than the true baseline evolution, so that this shift can be expected to deteriorate the test's performance. 
\par
\begin{figure}[!t]
\centering
\includegraphics[width=3.5in]{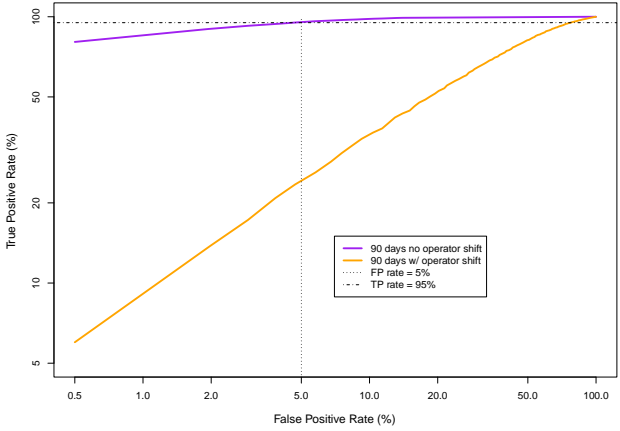}
\caption{ROC curves for the test applied to the original unshifted (purple) and operator-shifted (orange) baseline evolutions.  The time period shown here is days 0--90 in the fuel cycle.  The dotted vertical line corresponds to the FP rate of 5\%, while the dashed and dotted horizontal line corresponds to the TP rate of 95\%.}
\label{roc_opershift}
\end{figure}

Fig. \ref{roc_opershift} confirms this loss of sensitivity. Both ROC curves shown in this plot were obtained from the test using count rate data for days 0--90 in the cycle, assuming high counting statistics.  For this particular time period, the TP rate for the test applied to the shifted baseline was as low as one-ninth of that observed using the original baseline. For example, at the FP rate of 5\%, the TP rate of the former is 95\%, while that of the latter is 23\%.  In Section \ref{concl}, we discuss operational and experimental means to address the problem of deliberate misreporting.  \par
It should be noted that longer duration of data acquisition reduces the impact of malfeasance. As Table \ref{fp5perctable} shows, with high count rate data, the TP rates at 5\% FP rate are respectively 56\% and 99.99\% for 250 and 500 days of data acquisition. The complete ROC curves for the various acquisition times in the case of the shifted baseline are shown in Fig. \ref{roc_period}.  Hence, even in the presence of malfeasance,  the anomaly can be detected with high sensitivity if one acquires antineutrino data over the entire cycle.

\subsection{Effect of the Starting Point and Duration of the Data Acquisition Period}
\label{duration}
\begin{figure}[!t]
\centering
\includegraphics[width=3.5in]{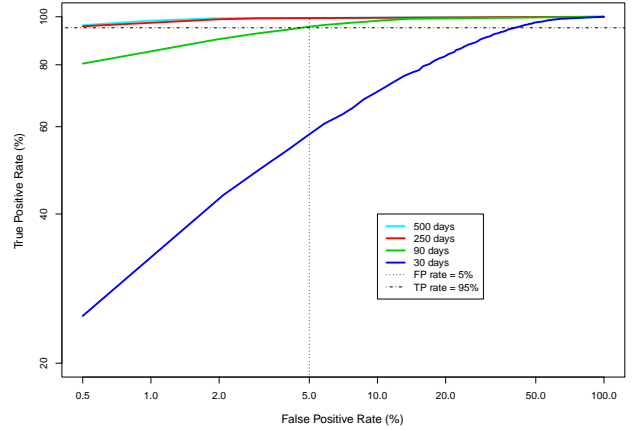}
\caption{ROC curves for the test using high count rates acquired over the full cycle, or 500 days (turquoise), days 0--250 of the cycle (blue), days 0--90 of the cycle (orange), and days 0--30 of the cycle (green). The dotted vertical line corresponds to the FP rate of 5\%, while the dashed and dotted horizontal line corresponds to the TP rate of 95\%.}
\label{roc_duration}
\end{figure}

Naturally, the estimates of the evolution coefficients $\hat{\gamma}^{(M)}_i$ and the test performance both improve as data are acquired for longer periods. In our ROC curve simulation, we considered the following four durations: days 0--500 (roughly full cycle length), days 0--250 (half cycle length), days 0--90 and days 0--30 in the cycle. Fig. \ref{roc_duration} shows the ROC curves for these four duration periods, assuming high count rates. At the FP rate of 5\%, the TP rate is 99.99\% for 500 days versus 99\%, 95\% and 58\% for 250, 90 and 30 days, respectively.
\par
Moreover, as Fig. \ref{counts_opershift} reveals, when the baseline is shifted due to incorrect input information, in addition to the duration of data acquisition, the location of the time window in the cycle during which the data are acquired will also affect the performance of the test.  For example, the shifted baseline evolution is less distinguishable from the  anomalous evolution in the first 250 days of the cycle than in the last 250 days.  The same is true when comparing the first 90 days to the last 90 days of the cycle. Therefore, we also compared the performance of the test for the shifted baseline using high count rate data from the first 90, last 90  (days 411-500), first 250, last 250 (days 251--500), and all 500 days of the cycle. 
\par
\begin{figure}[!t]
\centering
\includegraphics[width=3.5in]{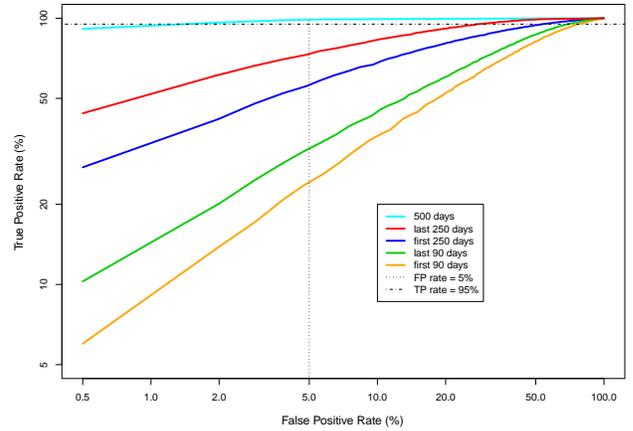}
\caption{ROC curves for the test applied to the operator-shifted baseline and high count rates acquired over the full cycle, or 500 days (turquoise), first 250 days of the cycle (blue), last 250 days of the cycle (red), first 90 days of the cycle (orange), and last 90 days of the cycle (green). The dotted vertical line corresponds to the FP rate of 5\%, while the dashed and dotted horizontal line corresponds to the TP rate of 95\%.}
\label{roc_period}
\end{figure}

Fig. \ref{roc_period} shows the ROC curves for these five periods for the case of the shifted baseline. As was noted earlier, as the number of days goes down, the test performance degrades. Moreover, the test applied to the count rate data for the last 250 days performs better than for the first 250 days because the shifted baseline and the anomalous evolutions are further apart at later times in the fuel cycle. The same is true when comparing the performance for the first 90 days to the last 90 days. However, the test is less sensitive to the starting point than to the duration of the data acquisition period. 
\par
These various effects are summarized in Table \ref{fp5perctable}. The effect of duration and period was very similar for the low count rates, so these results are not included in the table.    

\section{Impact on Detector Design and Operation} 
\label{detector}
The test performance described above can be used to guide the design of future safeguards antineutrino detectors. For a given anomalous scenario and desired true and false positive rate, a minimum antineutrino count rate requirement can be established. Within practical limits set by the reactor site, detector cost and complexity, a desired event rate may be achieved by adjusting the detector standoff distance, size or intrinsic efficiency. 

As discussed earlier, the antineutrino rate in the SONGS1 experiment \cite{SONGS1} was approximately 360 counts per day at beginning of cycle after subtraction of reactor-off backgrounds.  According to the ROC curve in Fig. \ref{roc_counts}, this antineutrino count rate  gives a 34\%  TP rate for a 5\% FP rate with a 90-day acquisition period. We assume that an acceptable test for IAEA safeguards or a similar monitoring regime will require at least  95\%  TP rate at the 5\% FP rate.  In the previous sections, we have shown that for the anomalous scenario we considered, a 2000 count per day net antineutrino event rate is necessary and sufficient to achieve this TP/FP rate combination. 

The SONGS1 detector was located  24.5 meters from the reactor core, with a 0.48 ton  target mass, and 11\% intrinsic detection efficiency  \cite{nuburnup}. An increase in event rate compared to SONGS1 could be accomplished by a combination of reduced standoff distance, increased detector target mass and/or increased intrinsic detection efficiency. For example, at 24.5 meter standoff, a one ton detector with 30\%  intrinsic efficiency, or a two ton detector with 15\% intrinsic efficiency would reach the 2000 count rate level and thus, the desired 95\%/5\% TP/FP rates. Alternatively, a one ton, 11\% efficient detector at 15 meter standoff would reach the same TP/FP rate combination. 

As shown in Table \ref{det_table}, previous antineutrino detectors had masses and efficiencies required to achieve the desired TP/FP rate performance. The  series of deployments at the Rovno reactor complex in the Ukraine is of particular interest since the efficiencies are high, while the overburden and other conditions are similar to those that would be encountered in many reactors under the IAEA safeguards. By contrast, the high efficiency of the CHOOZ detector reflects the state of the art for this class of detectors, but is achieved in part through significantly greater overburden and reduced ambient radioactivity  compared to the other experiments, so such a device is unlikely to be practical in a safeguards context.   
\begin{table}[h!b!p!]

\caption{Power, mass, standoff distance, efficiency, and signal-to-background ratios of some previous antineutrino experiments.}
\begin{tabular}{|l|l|l|l|l|l|l|}
\hline

Experiment 		 & Power     & Mass  	        &  Distance          & Efficiency   	   	&     Signal/Bkgd 	   \\ 
       	   			 & (GW)       &  (ton) 		&  \ \  \ (m) 	&  \ \ \ \ (\%)  	& 	\ \ \ \ \ Cts/Day		 	   \\ \hline
Rovno 1 \cite{Klimov}	 & 1.375	   &	\~0.5		& 18			& 20 		    		& 	909/149		    \\ 	\hline
Rovno 2 \cite{Korovkin} &1.375	   &	\~0.2		& 18			& 30 	    			&	267/94		    \\ \hline
CHOOZ \cite{Chooz}	  &	4.4	   &	5.0		& 1000		& 69.8  		 	&	24/1.2			    \\ \hline
Palo Verde \cite{PV}	 &      11.6	   &11.3		&800			&  10     			&	200/300   		    \\ \hline
SONGS1 \cite{SONGS1} &      3.4	   &0.64		&24.5		& 11 				&	564/105			    \\ \hline
Bugey \cite{Bugey}        &      3.4	   &0.64		&24.5		& 10	&	62/2.5						 \\ \hline
\end{tabular}
\label{det_table}
\end{table}

\section{Conclusions and Possible Future Work}
\label{concl}
This paper introduced a test procedure that determines whether a given antineutrino count rate evolution significantly deviates from that of the baseline.  The  procedure uses a quadratic model for the antineutrino count rate as a function of time since the beginning of the fuel cycle. However, the procedure can be adapted to a much wider class of models.  The procedure involves least squares estimation of the parameters in the quadratic model for the evolution in question and a multiple hypothesis testing procedure, known as False Discovery Rate (FDR), to determine whether at least one of the estimated parameters is significantly different from its baseline counterpart.  
\par

The anomalous operations identified in this paper do not constitute a diversion scenario per se, since we have not specified the ultimate fate of the removed fuel. Instead, we have estimated the sensitivity of antineutrino rate measurements to changes in typical civil power reactor fuel loadings. An important future exercise, best conducted by IAEA safeguards experts, is a fuller analysis of the reactor safeguards implications of this novel bulk accountancy method.

While the specific performance of the test will depend on the scenario, this work has identified the factors that most influence our ability to detect anomalous fuel loadings generally.  Among the factors that we considered, counting statistics, the presence of detector bias, and introduction of a systematic shift due to operator malfeasance had the most dramatic impact on the test performance. High counting statistics collected over longer periods of time in the absence of a deliberate shift in the baseline or detector bias yield the best performance and attain the target 95\% TP rate at the 5\% FP rate. We also found that the effect of a systematic error in detector bias response can be substantially reduced by an initial correction of the predicted to the observed count rates, or most effectively by an empirical calibration of detector response using antineutrino count rate data from a previous fuel cycle. The latter approach has the further advantage of lessening the dependence of the method on a reactor simulation. Changes in the starting point of data acquisition had a smaller impact on the performance. 
\par
Past experience has demonstrated that increasing the antineutrino count rate through efficiency or mass increases is achievable, so that our target 95\% TP / 5\% FP rate combination can be attained with practical detectors.   More problematic in a safeguards context is the issue of deliberate misreporting of power levels on the part of the operator that would undermine the statistical power of our test. While this is a serious concern, we note that the operator's misreporting must be fully consistent with the antineutrino data, which are independently acquired by and remain under the control of the safeguards inspector. This independently acquired information places an important additional constraint on the operator compared to current practice, in which declarations, along with item accountancy, are the primary sources of quantitative information about the reactor thermal power and fuel loading.  Moreover, the misrepresentation must be tuned to the particular anomalous operational state chosen by the operator. If different amounts or types of fissile material are removed, the hypothesis test may still detect a significant departure from the baseline. To further examine the robustness of this method, it is necessary to investigate a wider class of anomalous scenarios, varying both fuel and reactor type. 

As described in \cite{Huber}, a direct measurement of the antineutrino spectrum would provide sufficient information to simultaneously constrain both power and fissile isotopic content. This would severely undermine or even eliminate the benefit to the operator of misreporting the thermal power. However, since the antineutrino rate per energy bin will be necessarily reduced, the statistical power of the test may be compromised, or, alternatively, a larger detector may be required than is the case for a pure rate measurement.  In future work, we will apply a hypothesis testing procedure on a spectrally resolved antineutrino measurement,  including realistic statistical and systematic uncertainties, to quantify any additional sensitivity inherent in the spectral analysis.   
\par
Finally, as noted earlier,  we used an ORIGEN simulation of the SONGS Unit 2 reactor core. Assemblies were assumed to have no spatial extent:  the only spatial information in our calculation was the variation in distance of each pointlike assembly from the detector. A  full three-dimensional treatment of the assemblies would allow inclusion of effects, such as the variation of the centroid of fission over the cycle.

\section{Acknowledgments}
We thank the DOE Office of Nonproliferation Research and Engineering for their sustained support of this project. We also thank Nathaniel Bowden and Scott Kiff for insightful comments on an earlier version of this manuscript.  Finally, we express our gratitude to the management and staff of the San Onofre Nuclear Generating Station, for allowing us to deploy and take data with our prototype safeguards antineutrino detectors.

\newpage

\bibliographystyle{IEEEtran}
\bibliography{divdet}

\begin{IEEEbiography}{Vera Bulaevskaya and Adam Bernstein} are with the Lawrence Livermore National Laboratory in Livermore, CA.
\end{IEEEbiography}

\end{document}